\documentclass[%
reprint,
groupedaddress,
showpacs,
nofootinbib,
amsmath,amssymb,
aps,
prl,
superscriptaddress,
notitlepage,
floatfix
]{revtex4-2}

\usepackage{longtable}
\usepackage{graphicx}% Include figure files
\usepackage{dcolumn}% Align table columns on decimal point
\usepackage{bm}% bold math
\usepackage{color}
\usepackage{hyperref}% add hypertext capabilities
\definecolor{LinkColor}{rgb}{0,0,0.5}
\hypersetup{colorlinks=true,%
linkcolor=LinkColor,%
citecolor=LinkColor,%
filecolor=LinkColor,%
menucolor=LinkColor,%
urlcolor=LinkColor}
\usepackage[vertfit]{breakurl}
\usepackage[caption=false, labelformat=empty]{subfig}
\usepackage{microtype}
\usepackage{multirow}
\begin{document}
%\preprint{APS/123-QED}

\title{Direct measurement of the $^{13}$C$(\alpha$,n$)^{16}$O cross section into the $s$-process Gamow peak}% Force line breaks with \\

\author{
	G.F.\,Ciani
}
\affiliation{Gran Sasso Science Institute, Viale F. Crispi 7, 67100 L'Aquila, Italy}
\affiliation{INFN, Laboratori Nazionali del Gran Sasso (LNGS), 67100 Assergi, Italy}
\affiliation{Institute for Nuclear Research (ATOMKI), PO Box 51, HU-4001 Debrecen, Hungary}
\author{
    L.\,Csedreki
}
\affiliation{Gran Sasso Science Institute, Viale F. Crispi 7, 67100 L'Aquila, Italy}
\affiliation{INFN, Laboratori Nazionali del Gran Sasso (LNGS), 67100 Assergi, Italy}
\affiliation{Institute for Nuclear Research (ATOMKI), PO Box 51, HU-4001 Debrecen, Hungary}
\author{
    D.\,Rapagnani
}
\affiliation{Universit\`a di Napoli ``Federico II'', 80126 Napoli, Italy}
\affiliation{INFN, Sezione di Napoli, 80126 Napoli, Italy}
\author{
	M.\,Aliotta
}
\affiliation{SUPA, School of Physics and Astronomy, University of Edinburgh, EH9 3FD Edinburgh, United Kingdom}
\author{
    J.\,Balibrea-Correa
}
\affiliation{Universit\`a di Napoli ``Federico II'', 80126 Napoli, Italy}
\affiliation{INFN, Sezione di Napoli, 80126 Napoli, Italy}
\author{
    F.\,Barile
}
\affiliation{Universit\`a degli Studi di Bari, 70121 Bari, Italy}
\affiliation{INFN, Sezione di Bari, 70125 Bari, Italy}
\author{
	D.\,Bemmerer
}
\affiliation{Helmholtz-Zentrum Dresden-Rossendorf, Bautzner Landstr. 400, 01328 Dresden, Germany}
\author{
    A.\,Best
}
\email{Corresponding author: andreas.best@unina.it}
\affiliation{Universit\`a di Napoli ``Federico II'', 80126 Napoli, Italy}
\affiliation{INFN, Sezione di Napoli, 80126 Napoli, Italy}
\author{
	A.\,Boeltzig
}
\affiliation{Universit\`a di Napoli ``Federico II'', 80126 Napoli, Italy}
\affiliation{INFN, Sezione di Napoli, 80126 Napoli, Italy}
\author{
	C.\,Broggini
}
\affiliation{INFN, Sezione di Padova, Via F. Marzolo 8, 35131 Padova, Italy}
\author{
	C.G.\,Bruno
}
\affiliation{SUPA, School of Physics and Astronomy, University of Edinburgh, EH9 3FD Edinburgh, United Kingdom}
\author{
	A.\,Caciolli
}
\affiliation{INFN, Sezione di Padova, Via F. Marzolo 8, 35131 Padova, Italy}
\affiliation{Universit\`a degli Studi di Padova, Via F. Marzolo 8, 35131 Padova, Italy}

\author{
    F.\,Cavanna
}
\affiliation{INFN, Sezione di Torino, Via Pietro Giuria, 1, 10125 Torino, Italy}
\author{
    T.\,Chillery
}
\affiliation{SUPA, School of Physics and Astronomy, University of Edinburgh, EH9 3FD Edinburgh, United Kingdom}
\author{
    P.\,Colombetti
}
\affiliation{INFN, Sezione di Torino, Via Pietro Giuria, 1, 10125 Torino, Italy}
\author{
    P.\,Corvisiero
}
\affiliation{Universit\`a degli Studi di Genova, 16126 Genova, Italy}
\affiliation{INFN, Sezione di Genova, Via Dodecaneso 33, 16146 Genova, Italy}
\author{S.\,Cristallo}
\affiliation{INAF, Osservatorio Astronomico d'Abruzzo, Teramo, Italy}
\affiliation{INFN, Sezione of Perugia, Via A. Pascoli snc, 06123 Perugia, Italy}
\author{
    T.\,Davinson
}
\affiliation{SUPA, School of Physics and Astronomy, University of Edinburgh, EH9 3FD Edinburgh, United Kingdom}
\author{
	R.\,Depalo
}
\affiliation{Universit\`a degli Studi di Padova, Via F. Marzolo 8, 35131 Padova, Italy}
\affiliation{INFN, Sezione di Padova, Via F. Marzolo 8, 35131 Padova, Italy}
\author{
	A.\,Di Leva
}
\affiliation{Universit\`a di Napoli ``Federico II'', 80126 Napoli, Italy}
\affiliation{INFN, Sezione di Napoli, 80126 Napoli, Italy}
\author{
	Z.\,Elekes
}
\affiliation{Institute for Nuclear Research (ATOMKI), PO Box 51, HU-4001 Debrecen, Hungary}
\author{
    F.\,Ferraro
}
\affiliation{Universit\`a degli Studi di Genova, 16126 Genova, Italy}
\affiliation{INFN, Sezione di Genova, Via Dodecaneso 33, 16146 Genova, Italy}
\author{
    E.\,Fiore
}
\affiliation{Universit\`a degli Studi di Bari, 70121 Bari, Italy}
\affiliation{INFN, Sezione di Bari, 70125 Bari, Italy}
\author{
	A.\,Formicola
}
\email{Corresponding author: formicola@lngs.infn.it}
\affiliation{INFN, Laboratori Nazionali del Gran Sasso (LNGS), 67100 Assergi, Italy}
\author{
	Zs.\,F\"ul\"op
}
\affiliation{Institute for Nuclear Research (ATOMKI), PO Box 51, HU-4001 Debrecen, Hungary}
\author{
	G.\,Gervino
}
\affiliation{Universit\`a degli Studi di Torino, 10125 Torino, Italy} 
\affiliation{INFN, Sezione di Torino, Via P. Giuria 1, 10125 Torino, Italy}
\author{
	A.\,Guglielmetti
}
\affiliation{Universit\`a degli Studi di Milano, 20133 Milano, Italy}
\affiliation{INFN, Sezione di Milano, Via G. Celoria 16, 20133 Milano, Italy}
\author{
	C.\,Gustavino
}
\affiliation{INFN, Sezione di Roma La Sapienza, Piazzale A. Moro 2, 00185 Roma, Italy}
\author{
	Gy.\,Gy\"urky
}
\affiliation{Institute for Nuclear Research (ATOMKI), PO Box 51, HU-4001 Debrecen, Hungary}
\author{
	G.\,Imbriani
}
\affiliation{Universit\`a di Napoli ``Federico II'', 80126 Napoli, Italy}
\affiliation{INFN, Sezione di Napoli, 80126 Napoli, Italy}
\author{
    M.\,Junker
}
\affiliation{INFN, Laboratori Nazionali del Gran Sasso (LNGS), 67100 Assergi, Italy}
\author{
    M. Lugaro
    }
    \affiliation{Institute of Physics, ELTE Eötvös Loránd University, Budapest, Hungary}
    \affiliation{Konkoly Observatory, Research Centre for Astronomy and Earth Sciences, MTA Centre for Excellence, Budapest, Hungary}
\author{
    P. Marigo
    }
\affiliation{INFN, Sezione di Padova, Via F. Marzolo 8, 35131 Padova, Italy}
\affiliation{Universit\`a degli Studi di Padova, Via F. Marzolo 8, 35131 Padova, Italy}
\author{
    E. Masha
    }
\affiliation{Universit\`a degli Studi di Milano, 20133 Milano, Italy}
\affiliation{INFN, Sezione di Milano, Via G. Celoria 16, 20133 Milano, Italy}
\author{
	R.\,Menegazzo
}
\affiliation{INFN, Sezione di Padova, Via F. Marzolo 8, 35131 Padova, Italy}
\author{
	V.\,Mossa
}
\affiliation{INFN, Sezione di Bari, 70125 Bari, Italy}
\author{
	F.R.\,Pantaleo
}
\affiliation{Universit\`a degli Studi di Bari, 70121 Bari, Italy}
\affiliation{INFN, Sezione di Bari, 70125 Bari, Italy}
\author{
    V.\,Paticchio
}
\affiliation{INFN, Sezione di Bari, 70125 Bari, Italy}
\author{
    R.\,Perrino
}
\altaffiliation[Permanent address: ]{INFN Sezione di Lecce - Via Arnesano - 73100 Lecce, Italy}
\affiliation{INFN, Sezione di Bari, 70125 Bari, Italy}
\author{
	D.\,Piatti
}
\affiliation{INFN, Sezione di Padova, Via F. Marzolo 8, 35131 Padova, Italy}
\affiliation{Universit\`a degli Studi di Padova, Via F. Marzolo 8, 35131 Padova, Italy}
\author{
    P.\,Prati
}
\affiliation{Universit\`a degli Studi di Genova, 16126 Genova, Italy}
\affiliation{INFN, Sezione di Genova, Via Dodecaneso 33, 16146 Genova, Italy}
\author{
    L.\,Schiavulli
}
\affiliation{Universit\`a degli Studi di Bari, 70121 Bari, Italy}
\affiliation{INFN, Sezione di Bari, 70125 Bari, Italy}
\author{
    K.\,St\"ockel
}
\affiliation{Helmholtz-Zentrum Dresden-Rossendorf, Bautzner Landstr. 400, 01328 Dresden, Germany}
\affiliation{Technische Universit\"at Dresden, Institut f\"ur Kern- und Teilchenphysik, Zellescher Weg 19, 01069 Dresden, Germany}

\author{
	O.\,Straniero
}
\affiliation{INAF, Osservatorio Astronomico d'Abruzzo, Teramo, Italy}
\affiliation{INFN, Laboratori Nazionali del Gran Sasso (LNGS), 67100 Assergi, Italy}
\author{
	T.\,Sz\"ucs
}
\affiliation{Institute for Nuclear Research (ATOMKI), PO Box 51, HU-4001 Debrecen, Hungary}
\author{
	M.P.\,Tak\'acs
}
\affiliation{Helmholtz-Zentrum Dresden-Rossendorf, Bautzner Landstr. 400, 01328 Dresden, Germany}
\affiliation{Technische Universit\"at Dresden, Institut f\"ur Kern- und Teilchenphysik, Zellescher Weg 19, 01069 Dresden, Germany}

\author{
    F.\,Terrasi
}
\affiliation{Universit\`a degli Studi della Campania ``L. Vanvitelli'' (Caserta), 81100 Caserta, Italy}
\affiliation{INFN, Sezione di Napoli, 80126 Napoli, Italy}
\author{D.\,Vescovi}
\affiliation{INFN, Sezione of Perugia, Via A. Pascoli snc, 06123 Perugia, Italy}
\affiliation{Goethe University, Max-von-Laue-Strasse 1, Frankfurt am Main 60438, Germany} 
\author{
    S.\,Zavatarelli
}
\affiliation{INFN, Sezione di Genova, Via Dodecaneso 33, 16146 Genova, Italy}

\collaboration{LUNA Collaboration}\noaffiliation

\date{\today}% It is always \today, today,

\begin{abstract}
One of the main neutron sources for the astrophysical $s$-process is the reaction $^{13}$C$(\alpha$,n$)^{16}$O, taking place in thermally pulsing Asymptotic Giant Branch stars at temperatures around 90\,MK.
To model the nucleosynthesis during this process the reaction cross section needs to be known in the 150-230\,keV energy window (Gamow peak).
At these sub-Coulomb energies cross section direct measurements are severely affected by the low event rate, making us rely on input from indirect methods and extrapolations from higher-energy direct data.
This leads to an uncertainty in the cross section at the relevant energies too high to reliably constrain the nuclear physics input to $s$-process calculations.
We present the results from a new deep-underground measurement of $^{13}$C$(\alpha$,n$)^{16}$O, covering the energy range 230-300\,keV, with drastically reduced uncertainties over previous measurements and for the first time providing data directly inside the $s$-process Gamow peak. Selected stellar models have been computed to estimate the impact of our revised reaction rate. For stars of nearly solar composition, we find sizeable variations of some isotopes, whose production is influenced by the activation of close-by branching points that are sensitive to the neutron density, in particular the two radioactive nuclei $^{60}$Fe and $^{205}$Pb, as well as $^{152}$Gd.  
\end{abstract}


\maketitle
%\linenumbers
Low-mass Asymptotic Giant Branch (AGB) stars are major production sites of heavy elements in the Universe (for a recent review see \cite{Sneden:2008}). Their interior contains a carbon-oxygen core surrounded by a thin He-rich mantel and a H-rich envelope. Periodically, these stars undergo thermonuclear instabilities, He-burning flashes, called thermal pulses (TPs). Each He-flash generates a large convective zone that mixes the C produced by the triple-$\alpha$ reaction up to the top of the He mantel. Later on, the shell-H burning, always active at the base of the envelope, dies down and, in turn, the external convection penetrates the He-rich mantel.  
This phenomenon, which moves the nucleosynthesis yields up to the stellar surface, is called the \emph{third dredge up}. As early recognized, the creation of a $^{13}$C-pocket within the He-rich mantel, through the reaction $^{12}$C(p,$\gamma)^{13}$N$(\beta^+)^{13}$C, is a byproduct of these recursive mixing episodes \cite{Straniero:2006,gallino:1998}. Such a thin pocket (a few $10^{-5}$M$_\odot$ of $^{13}$C) harbors one of the most important nucleosynthesis sites in the Universe.  
During the period between two TPs, the temperature attains about 90\,MK and $^{13}$C is activated as a neutron source through the reaction $^{13}$C$(\alpha$,n$)^{16}$O.
This process provides a relatively slow neutron flux ($\approx 10^{7}$\,neutrons/s/cm$^2$) for about $10^4$ years each time. Starting from seed nuclei in the iron region, this neutron flux slowly builds up heavy elements along the line of stability \cite{Kaeppeler:2011}.
This $s$-process (\emph{slow-neutron-capture}) is responsible for the production of about half of all the heavy elements (A$\geq 90$) in the Universe.

In order to constrain this important nucleosynthesis process, the cross section of the $^{13}$C$(\alpha$,n$)^{16}$O neutron source needs to be known in the astrophysical energy window (the Gamow peak) around $E_{0}$= 150-230 keV\footnote{All energies in this manuscript are center-of-mass energies, unless explicitly stated otherwise.}.
%Direct measurements at these energies are very challenging, as the cross section is strongly suppressed by the Coulomb barrier.
The available direct cross section data in the lower energy interval $280<E<350$\,keV  \cite{Drotleff:1993,Heil:2008a} are affected by uncertainties $\geq 40\%$.
%Direct cross section data  so far extend only down to 280 keV, with very large uncertainties ($> 40\%$) below 350 keV\cite{Drotleff:1993,Heil:2008a}.
Any effort of pushing direct reaction measurements to lower energies is basically rendered futile by the steep drop of the cross section and the presence of natural and instrumental backgrounds.
Extrapolation of the cross section into the Gamow peak is further complicated by a broad state at $E_{x} = 6356 (8)$\,keV ($J^{\pi}=\frac{1}{2}^{+}$) just near the $\alpha$-threshold in $^{17}$O (S$_{\alpha} = 6359$\,keV \cite{Tilley1993}).
%The exact energy of this state is reevaluated quite frequently, some measurements place it above threshold (e.g., ref. \cite{Faesterman15}), some below. The influence of the energy on the uncertainty budget is discussed in the analysis part of this manuscript.
The cross section evaluation in the Gamow peak requires a careful matching between the tail of this near-threshold state and the higher-energy experimental data.
While this state has been the focus of great experimental attention over the past years \cite{Kubono2003,Johnson2006,Avila2015,Trippella2017,Keeley2018} its influence remains a major source of uncertainty for the $s$-process \cite{DeBoer:2020} and the need for more cross section data to fill the gap has been  frequently expressed \cite{Cristallo:2018, DeBoer:2020}.

In order to provide direct data at low energies to better constrain the cross section inside the Gamow peak the LUNA collaboration has performed an intensive experimental campaign at the deep-underground accelerator LUNA400 \cite{Formicola:2003} inside the Gran Sasso National Laboratory (LNGS).
The LNGS neutron background flux ($\approx 10^{-7}$n/cm$^2$/s) \cite{Best:2016}) is dominated by the natural radioactivity of the surrounding rock and it is up to four orders of magnitude lower than on the surface of the Earth, improving the sensitivity over previous studies.

The experimental setup and the target analysis are  described in detail in refs. \cite{Ciani:2020,CSEDREKI2021165081,CianiPhd}, following is a brief summary. The accelerator provided a He$^+$ beam on target of up to 150 $\mu$A with $\alpha$ energies of 305-400\,keV, corresponding to c.m. energies in the range 233-306\,keV. Deposited charges vary from 15\,C at the highest to 90\,C at the lowest energy.
The beam impinged upon water-cooled targets made of an 99\% enriched $^{13}$C evaporated on a 0.2\,mm thick Ta backing \cite{Ciani:2020}.

Near the target, the beam passed a liquid nitrogen cooled shroud  and an electrically insulated collimator at a negative voltage of 300\,V (to suppress the effects of secondary electrons).
%The latter part of the target chamber was electrically isolated and served as a Faraday cup for beam intensity measurement.
%A liquid-nitrogen cooled cold finger at -300 V potential extended close to the front face of the target to both improve the vacuum inside the chamber (to the 10$^{-7}$ mbar range) and to reflect back sputtered electrons onto the target.
The neutrons produced by the $^{13}$C($\alpha$,n)$^{16}$O reaction, with energies around 2.4\,MeV, were thermalized in a polyethylene moderator and detected by 18 $^3$He-filled proportional counters with stainless steel housing. 
%\footnote{The counters were positioned in two rings around the beam axis, with 6 counters in the inner and 12 counters in the outer ring.
%The moderator was surrounded by 5 or 10 cm of 5\% borated polyethylene shielding, depending on the energy of the measurement.}.
Two geometrical detector configurations (a vertical and a horizontal orientation) were used to optimize  the absolute neutron detection efficiency, $34 \pm 3$\% and $38 \pm 3$\%, respectively \cite{CSEDREKI2021165081}. Moreover, the measurement reproducibility was checked by separately analysing datasets acquired with the two detector configurations at the same energy, finding agreement in the final results. 
%The energy ranges covered using the two setups overlapped with each other in order to control possible systematic differences in their efficiency calibrations.
The detector signals were digitized with a 100 MHz sampling rate and analyzed offline to suppress the internal background from $\alpha$-decays in the steel housing of the counters using a custom pulse-shape discrimination (PSD) technique \cite{Balibrea:2018}.
The PSD in combination with the borated polyethylene shielding (5 and 10 inches, depending on the setup) around the detector as well as the underground location results in a total background rate inside the neutron signal region of $1.2 \pm 0.1$\,counts/hour for both configurations, %$3.34 \pm 0.11$ (vertical) and $3.08 \pm 0.09$ (horizontal) counts/hour ( check the number!!!), %$1.2 \pm 0.1$ counts/hour.
leading to an improvement by more than two orders of magnitude compared with previous experiments.

Blank backings were irradiated obtaining a background level of 1.3$\pm$\,0.2  counts/hour, in agreement with the environmental one. The conclusion was that beam induced background was negligible. The evaporated $^{13}$C targets were characterized in term of homogeneity and thickness immediately after the evaporation at ATOMKI by means of the $^{13}$C(p,$\gamma)^{14}$N resonance at $E_{\rm {r,lab}}$= 1748\,keV. The average target thickness at resonance energy was 5\,keV, corresponding to 170\,nm. Because of the cross section reduction of one order of magnitude, the differential neutron yield becomes negligible from reactions occurring beyond 150\,nm inside the target, so all the targets can be considered to be of the same effective thickness.
At LUNA, target quality was frequently checked (every 1.5\,C accumulated charge) using direct $\gamma$-ray measurements of the $^{13}$C(p,$\gamma)^{14}$N reaction \cite{Ciani:2020}.

The cross section was calculated as follows:
\begin{equation}
\label{eq:yield}
Y=\frac{\eta(E_{\alpha})}{Qe}\int^{E_{\rm \alpha}}_{E_{\alpha}-\Delta E}\frac{\sigma(E)}{\epsilon_{\rm{eff}}(E)}dE \quad ,
\end{equation}
where $Y$ is the number of detected neutrons per projectile (PSD corrected), $\eta$ the neutron detection efficiency, $Qe$ is the incident number of particles on the target, $E_{\alpha}$ the beam energy, $\epsilon_{\rm{eff}}(E)$ the effective stopping power and $\Delta E$ the projectile energy loss in the target. 

%For measurements of the $\alpha$-induced reaction cross sections,
The maximum accumulated $\alpha$-charge on each target was limited to 3\,C, corresponding to at most a 30\% degradation.% due to stoichiometry modification.
The lowest energies $E=245$ and $E=233$\,keV required special attention, as the statistics collected during a single run was insufficient to obtain a reliable estimate of the cross section.
Two independent approaches were used in the analysis. In the first one, all runs at the same energy with similar target degradation levels were summed together and the cross sections for each ``subset'' were calculated and combined to the cross section for this energy.In the second one, the two lowest energies were also analyzed using a Bayesian approach \cite{Caldwell:2009}, see the supplemental material for details. 

In Figure \ref{fig:comparisoncross}, the 233\,keV cross sections as a function of target degradation are compared with the Bayesian results. The mean values are in agreement; the ``grouping'' method was used for the extraction of the cross sections presented later in this work.% The procedure and results are presented in more detail in the supplemental material to this Letter.
\begin{figure}[tb]
    \centering
    \includegraphics[width=\columnwidth]{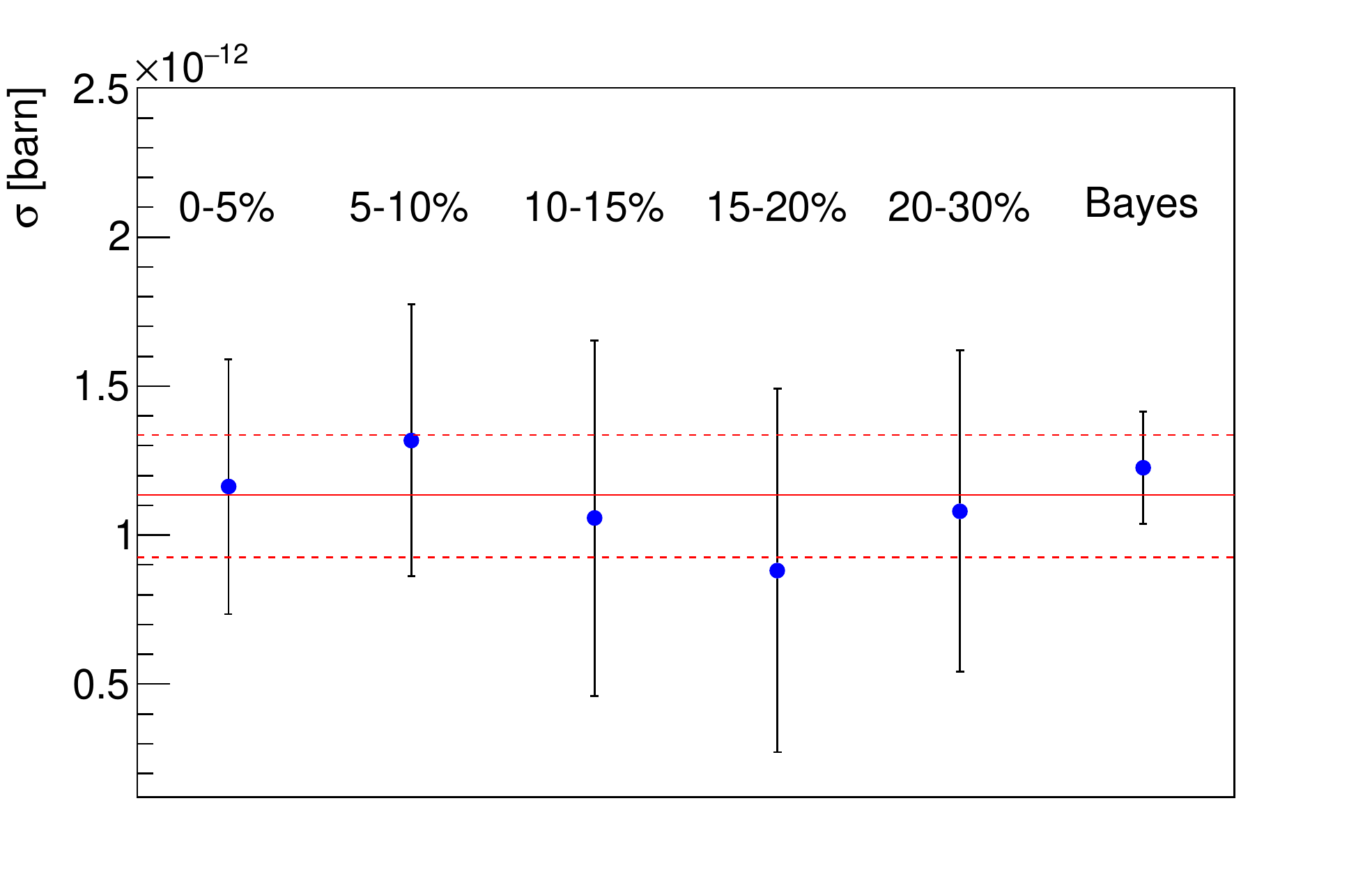}
    \caption{Comparison of the cross section at 233\,keV for different target degradation and the Bayesian results. The solid and dashed lines represent the weighted average of the groups and the corresponding uncertainty, respectively.}
    \label{fig:comparisoncross}
\end{figure}

The experimental results are summarized in Table \ref{tab:results}, where we also show the $S$-factor\footnote{The astrophysical $S$-factor is defined as $\sigma(E)=S(E)\frac{1}{E}e^{-2\pi\eta}$ and is commonly used to remove the strong energy dependence in the cross section due to the Coulomb repulsion. $\eta = 0.1575Z_aZ_b\sqrt{\frac{1}{E}\frac{M_aM_b}{M_a+M_b}}$ is the Sommerfeld parameter.} after correcting for the electron screening effect\footnote{A screening potential of 937\,V was used for the calculations: the corresponding correction is less then 10\% for all the energies.} (bare in Table \ref{tab:results}), following refs. \cite{Strieder:2012, Bracci:1990, Huang:1976}. The Table includes statistical and systematic uncertainties. The latter are: 3\% for the charge integration, 5\% for stopping power calculated with SRIM-2010 \cite{Ziegler:2010} and about 8\% for the detection efficiency. The uncertainty of the beam energy is around 300\,eV, and the beam energy spread is less than 100\,eV\ \cite{Formicola:2003}.

\begin{table}[tb]
    \caption{Experimental and electron screening-corrected $S$-factors. The data format is data$\pm$stat$\pm$syst, where syst is a common systematic uncertainty of 10 $\%$ (further details in the text).}
\begin{ruledtabular}
\begin{tabular}{ccc}
    Energy    & $S$-factor          &  $S$-factor      \\\relax
      [keV]       & [10$^5$MeV\,b]     & [10$^5$MeV\,b]  \\
                & &             (bare)    \\
\hline
306 &  8.06 $\pm$ 0.18 $\pm$ 0.8 & 7.61 $\pm$ 0.17 $\pm$ 0.8  \\
298 &  8.1  $\pm$ 0.3 $\pm$ 0.8  & 7.6  $\pm$ 0.3 $\pm$ 0.8 \\
291 &  7.3  $\pm$ 0.3 $\pm$ 0.7 & 6.8  $\pm$ 0.3 $\pm$ 0.7 \\
283 &  8.6  $\pm$ 0.3 $\pm$ 0.9 & 8.0  $\pm$ 0.3 $\pm$ 0.8 \\
275 &  9.2  $\pm$ 0.6 $\pm$ 0.9 & 8.6  $\pm$ 0.5 $\pm$ 0.9 \\
260 &  8.7  $\pm$ 0.8 $\pm$ 0.9 & 8.1  $\pm$ 0.7 $\pm$ 0.8 \\
245 &  11.7 $\pm$ 1.7 $\pm$ 1.2 & 10.8 $\pm$ 1.5 $\pm$ 1.1\\
233 &  12.7  $\pm$ 2.3 $\pm$ 1.3 & 11.6  $\pm$ 2.1 $\pm$ 1.2\\
\end{tabular}
\end{ruledtabular}
\label{tab:results}
\end{table}

\begin{figure}[tb]
    \centering
    \subfloat{\includegraphics[width=\columnwidth]{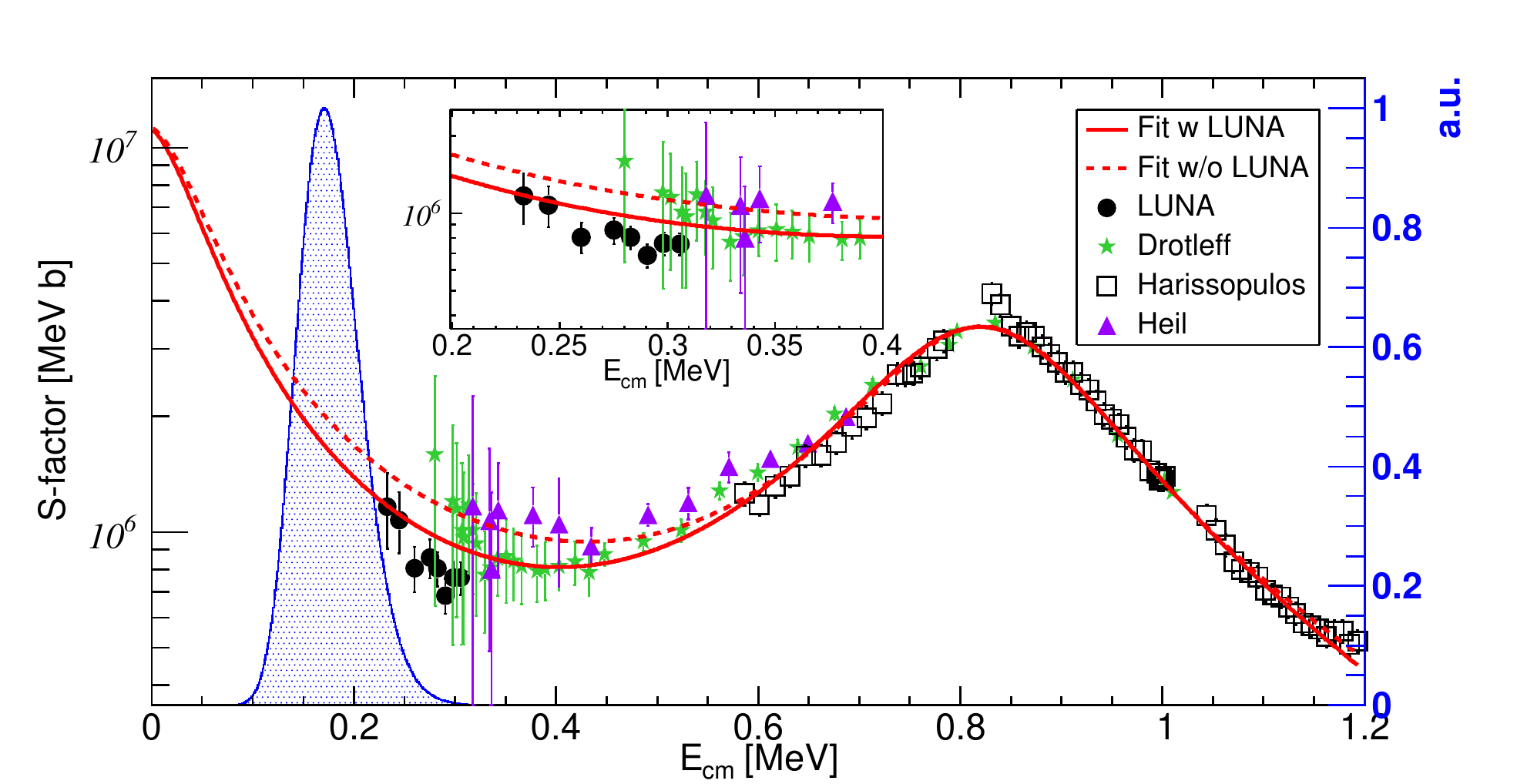}}\\
    \subfloat{\includegraphics[width=\columnwidth]{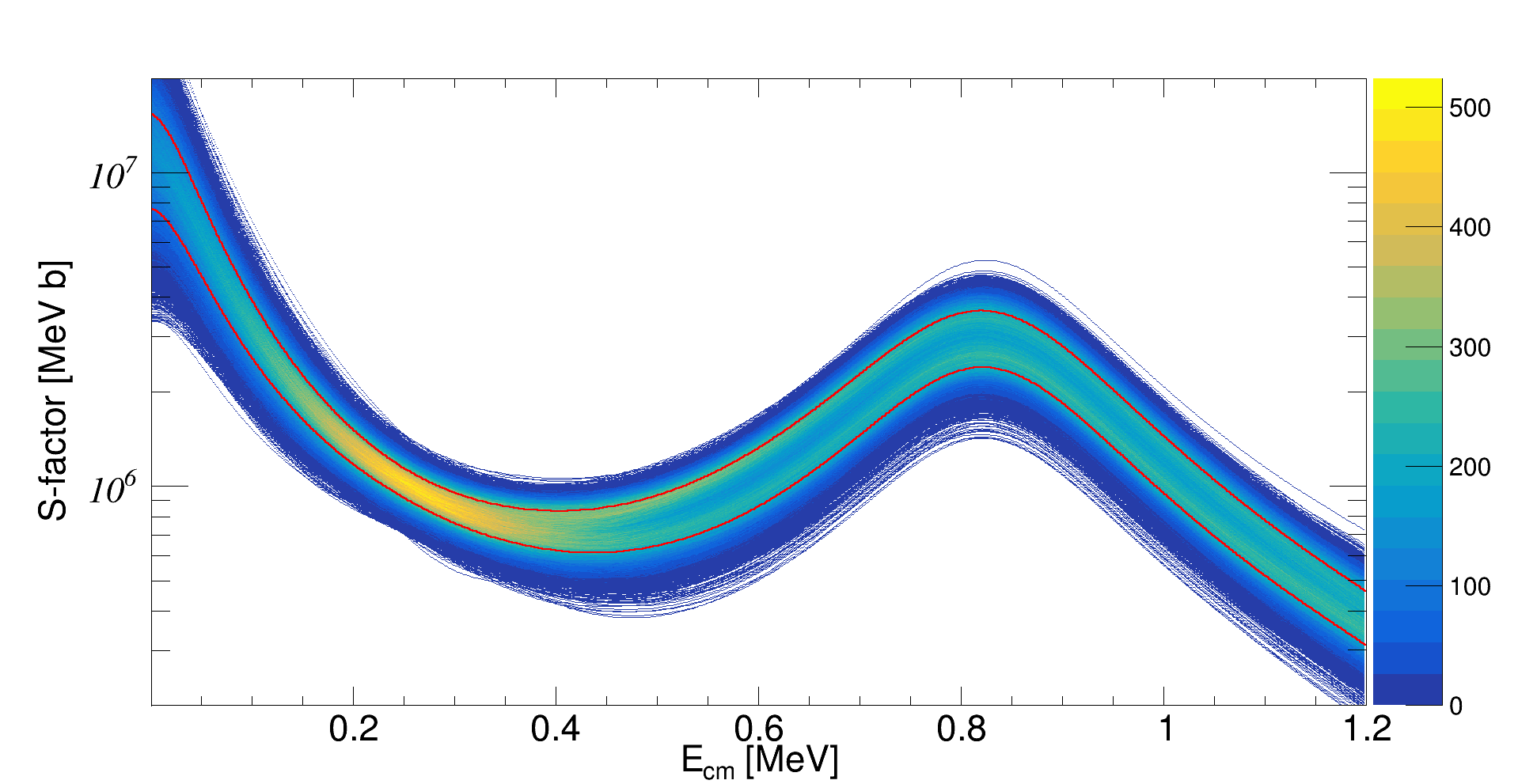}}
    \caption{Top: Astrophysical $S$-factor of $^{13}$C$(\alpha$,n$)^{16}$O. The lines show the results of two R-matrix analyses, with and without the new data. The Harissopulos dataset was normalized according to value in Table \ref{tab:rmatrix}. The blue curve is the ``Gamow peak" at 90\,MK and the right y axis refers to its relative scale. Bottom: Monte Carlo probability density for the $S$-factor.}
    \label{fig:sfactor}
\end{figure}

The new data extend the energy range covered by direct cross-section measurements into the $s$-process Gamow peak, but an extrapolation towards zero energy, also taking into account the near-threshold state ($E_x = 6356$\,keV, J$^{\pi} = \frac{1}{2}^+$), is still required. This was done with an R-matrix analysis using the code Azure2 \cite{Azuma:2010}. The low-energy cross section is dominated by two broad states, the already mentioned 
near-threshold state and a $\frac{3}{2}^+$, $E_x = 7215$\,keV one. Narrow resonances in the energy range covered by the analysis ($E<1.2$\,MeV) do only have very localized effects on the cross sections and were omitted. As the threshold state was assumed to be $\alpha$-bound, an asymptotic normalization coefficient instead of an $\alpha$ partial width was used \cite{Avila2015}. Energy and width or asymptotic normalization constant (ANC) of the threshold state were kept fixed (note that they were varied in the Monte Carlo analysis, see below). Channel radii of 4.15\,fm and 6.684\,fm were used for the neutron and $\alpha$ channels, respectively. In addition to the cross sections from Heil and Drotleff, the data by Harissopulos \cite{Harrisopulos:2005} were included in the analysis, as they cover a wider energy range and help better constrain the $\approx$ 800\,keV resonance. A normalization factor of 1.37 was applied to the latter to match the absolute scales of the different data sets. Two high-energy poles were included to take into account the influence of higher-lying resonances on the cross section. Their widths and that of the $E_x$ = 7215\,keV state were kept free. The parameters resulting from the R-matrix fit are given in Table \ref{tab:rmatrix}. The top panel of Figure \ref{fig:sfactor} shows the experimental $S$ factors and the R-matrix fits with (solid line) and without (dashed line) our new LUNA data.

\begin{table}[tb]
\caption{R-matrix parameters and data sets.}
\begin{ruledtabular}
\begin{tabular}{ccc|cc}
    $E_x$  &  $\Gamma_n$ & $\Gamma_{\alpha}|$ANC & \multirow{2}{*}{Dataset} & \multirow{2}{*}{Norm.} \\
 $[$keV$]$ & $[$keV$]$ & $[$keV]$|$[fm$^{-1/2}$ $]$ & & \\
\hline
6356 & 124 & $5.44 \times 10^{90}$ & Drotleff \cite{Drotleff:1993} & 1 \\
7215 & 305.3  & $9.75 \times 10^{-2}$ & Heil \cite{Heil:2008a} & 1 \\
15000 & $2.42 \times 10^{4}$ & $6.04 \times 10^{5}$ & Harissopulos \cite{Harrisopulos:2005} & 1.37 \\
15000 & $4.33 \times 10^{2}$ & $6.02 \times 10^{5}$ & This work & 1 \\
\end{tabular}
\end{ruledtabular}
\label{tab:rmatrix}
\end{table}

The uncertainties in the final cross section were investigated using a Monte Carlo (MC) approach. The ANC and the neutron partial width of the threshold state, as well as the absolute scales of the four experimental data sets, were randomly sampled from a Gaussian distribution according to their respective uncertainties. To be conservative, a 12\% uncertainty was assumed for the absolute scale of each data set. 30000 MC variations were evaluated and the R-matrix cross sections of each run were saved for later processing. The density map of the results and traditional 1 $\sigma$ contours are displayed in Figure \ref{fig:sfactor}. 
As mentioned above, the cross sections from \cite{Harrisopulos:2005} are rescaled to match the \cite{Drotleff:1993} and \cite{Heil:2008a} data, as in other recent papers. However, there are no strong motivations for doing so, and one could choose to instead base the normalization on the Harissopulos et al. data (as suggested by a recent measurement \cite{Febbraro:2020}). To investigate the effect of the two different normalizations we performed R-matrix calculation using data by Harissopulos et al. as a reference for the normalization of \cite{Drotleff:1993} and  \cite{Heil:2008a}. Inside the Gamow peak, the effect is only of the order of 5\%, increasing towards higher energies.
The absolute scale of the normalization of the historical data is still a matter for debate.
Therefore in the MC procedure, only for the sake of the determination of a lower limit, we also considered the case (for half of the total trials) of using unscaled \citet{Harrisopulos:2005} data, while the \cite{Drotleff:1993} and \cite{Heil:2008a} normalizations were changed accordingly. It is worth noting that this additional source of uncertainty contributes only marginally, about 5\%, at the $s$-process energies, where the cross section is well constrained by the present data, while it has a larger impact at higher energies\footnote{New measurements of the higher-energy cross section are planned at various facilities worldwide, including with the upcoming LUNA MV accelerator at the LNGS\cite{Ferraro2021Fr}, addressing this uncertainty.}.
As will be discussed below, the establishment of a reliable lower limit is crucial for the determination of possible nucleosynthesis variations.
In total, three different rates were calculated: ``LUNA", using the R-matrix best fit including the new cross sections, ``no-LUNA", the best fit without the new data, and ``low-LUNA" using the 5th percentile of the fit adopting the Harissopulos normalization. The ``no-LUNA", and ``low-LUNA" cover approximately the $\pm$95\% percentiles around the best fit.

Finally, the astrophysical reaction rate $R = N_A\langle\sigma v\rangle$ as a function of stellar temperature was calculated (in units of $\text{cm}^3 \text{mol}^{-1} \text{s}^{-1}$) by integration of the R-matrix cross section:
\begin{equation}
    \label{eq:rate}
    R = \frac{3.7318}{T_9^{3/2}} \sqrt{\frac{M_0 + M_1}{M_0 M_1}} \times \int_0^{\infty} E \sigma(E) e^{-11.605 E/T_9} dE \enspace,
\end{equation}
where $T_9$ is the temperature in GK, the energy is given in MeV, the cross section $\sigma$ in b and the $M_i$ are the atomic masses of the reaction partners. Probability density functions for the rate were generated based on the results of the various MC cross sections. The contributions from the narrow resonances were as usual summed to the final rate. Tabulated results are shown in the supplemental material accompanying this Letter.

%--------- evaluation of the astrophysical impact
In the $s$-process in AGB stars \cite{straniero:1995,gallino:1998}, shortly after a third-dredge-up episode, a $^{13}$C-pocket forms within the He-rich mantel. During the following interpulse period, this region heats up and, around $\sim$ 80-100\,MK, the $^{13}$C$(\alpha$,n$)^{16}$O reaction starts to release neutrons. The high neutron exposure ($\sim0.4$\,mb$^{-1}$ in solar metallicity stars) coupled to a low neutron density (a few $10^6$ n/cm$^3$) are the two major features of the resulting $s$-process nucleosynthesis. The first  ensures a neutron flux over a time long enough to produce a large overabundance of all the elements belonging to the main component of the $s$-process ($A> 90$), while the second favors  $\beta^-$ decays over neutron captures at the various branching points of the $s$-process path. In most cases, the $^{13}$C-pocket is fully consumed during the interpulse period. However, if a small amount of $^{13}$C survives (i.e., if the reaction rate is low enough), it will be engulfed into the convective shell powered by the subsequent thermal pulse and burned at a higher temperature ($\sim 200$\,MK) \cite{cristallo:2009}. 
This second (convective) neutron burst is characterized by a higher neutron density ($>10^{9}$ n/cm$^3$) but much lower neutron exposure, than the main (radiative) event. As a consequence, it does not modify the bulk of stellar yields, but may affect some key isotopes at the $s$-process branching points \cite{bisterzo:2015}. Extant models of low-mass AGB stars  have shown that this second neutron burst may occur during the first few thermal pulses, in stars with metallicity Z$\geq 0.01$ and that its efficiency is sensitive to the adopted $^{13}$C$(\alpha$,n$)^{16}$O reaction rate \cite{Cristallo:2018}.

In order to evaluate the impact of the new reaction rate on the $s$-process, we have calculated three models of an AGB star with mass M=2 M$_\odot$, metallicity Z$=0.02$ and Y$=0.27$, under the three different assumptions (``LUNA" and the $\sim \pm$ 95 \% values corresponding to ``no-LUNA" and ``low-LUNA") for the $^{13}$C$(\alpha$,n$)^{16}$O rate mentioned above. In all the three models, some $^{13}$C survives at the end of the first two interpulse periods and is burned at high temperature in the convective thermal pulse. Stronger effects of this second neutron burst are expected for the low-LUNA case. The results are compared in Figure \ref{fig:difx}. More details on the stellar models are given in the supplemental material, which includes Refs. \cite{Cristallo:2011, Vescovi:2020, Karakas:2016}.

\begin{figure}[tb]
    \centering
    \includegraphics[width=\columnwidth]{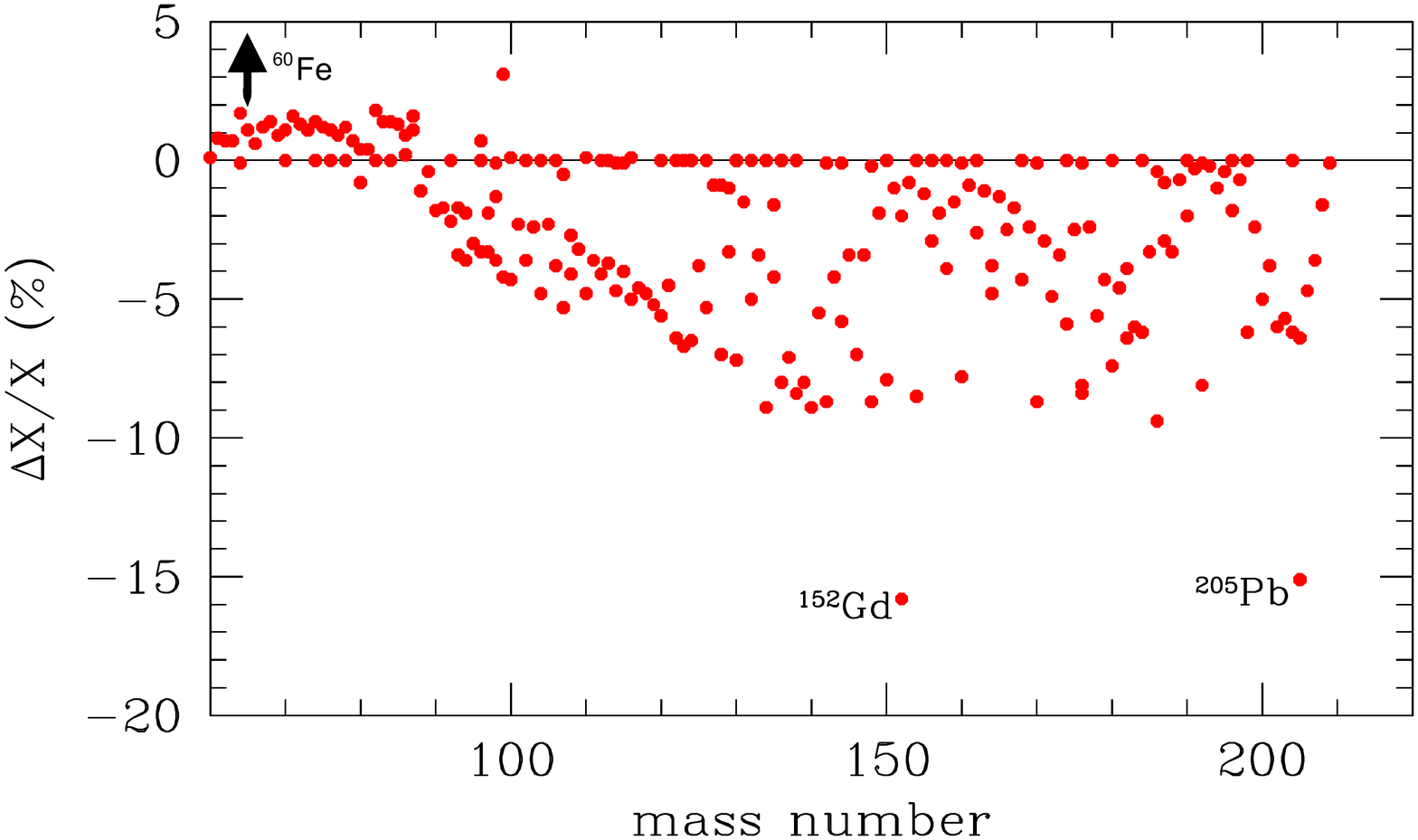}
    \caption{Percentage variations %at the AGB tip of the surface
    of the mass fractions of heavy isotopes in the low-LUNA model with respect to the NO-LUNA model. In both models, the final surface composition is the cumulative result of 11 third dredge up episodes. The $^{60}$Fe variation (+102\%) is out of the scale and marked by an arrow.}
    \label{fig:difx}
\end{figure}

Most of the nuclei belonging to the main component are depressed when the $^{13}$C$(\alpha$,n$)^{16}$O rate is lowered. This is the natural consequence of the suppression of the radiative $s$-process. The reduction of the surface abundances is stronger for the heavier isotopes ($A>130$). The heavy-s (Ba, La, Ce, Nd, Sm) to light-s (Sr, Y, Zr) abundance ratio is an important spectroscopic index, often used to probe the mean neutrons to seeds ratio of an $s$-process site \cite{Busso:1999}. At the end of the AGB phase, the [heavy-s/light-s] ratio of the low-LUNA model is $\sim6\%$ smaller than in the no-LUNA model. The abundance variations are generally small ($\leq 10\%$), with interesting exceptions, i.e., $^{60}$Fe, $^{152}$Gd and $^{205}$Pb. All these isotopes are sensitive to the neutron density because of the existence of close-by branching points. $^{152}$Gd cannot be produced by the r-process and, except for a small p-process contribution, is mainly synthesized by the $s$-process. On the other hand, $^{60}$Fe and $^{205}$Pb are short-lived radioactive isotopes that were found to be alive in the early solar system \cite{wasserburg2006}. The $^{60}$Fe is produced when the neutron density is high enough to allow neutron captures at the $^{59}$Fe branching point  (half-life 44.5\,d). Therefore, its final abundance is enhanced in case of the activation of the second (convective) neutron burst. With respect to the no-LUNA model, we find that the $^{60}$Fe final mass fraction is a 30\% higher, in the LUNA model, and a factor of 2 higher, in the low-LUNA model. On the contrary, the production of both $^{152}$Gd, and $^{205}$Pb requires low neutron density, while they are mainly destroyed in case of high neutron density. Indeed, the first isotope is bypassed by the $s$-process when the $^{151}$Sm may capture a neutron before decaying into $^{151}$Eu. Similarly, the $^{205}$Pb production is suppressed when the $^{204}$Tl branch is open. Therefore, their production is reduced when the $^{13}$C$(\alpha$,n$)^{16}$O rate is lowered, as it happens for all the other isotopes of the main component, and a further decrease of their abundances occurs in case of a more efficient second (convective) neutron burst. As a result, the final mass fractions of both isotopes are reduced by 15\% at the 95\% lower bound. 

To conclude, the present work reports a much improved calculation of the $^{13}$C$(\alpha$,n$)^{16}$O reaction rate at $T\sim 90$\,MK, for the first time based on direct data inside of the Gamow window. The reduced uncertainty will help our understanding of the $s$-process branching points that are sensitive to the neutron density. We find that the new low-energy cross-section measurements imply sizeable variations of the $^{60}$Fe, $^{152}$Gd and $^{205}$Pb yields. Other isotopes, whose production or destruction are influenced by close-by branching points, such as the two neutron-magic nuclei $^{86}$Kr and $^{87}$Rb as well as $^{96}$Zr, are less affected by a variation of the $^{13}$C$(\alpha$,n$)^{16}$O reaction rate,  mainly because of their higher initial (solar) abundance. However, we cannot exclude that larger changes may occur in models with different initial mass and composition. For this reason, a more extended set of AGB models is required to accurately evaluate the general impact on the galactic chemical evolution.

\begin{acknowledgments}
This work has been supported by the INFN. D. Ciccotti of the LNGS proved to be invaluable over the  course of the campaign. We acknowledge the support of the mechanical workshops of INFN-LNGS and INFN-Na. Financial support is acknowledged as follows. D. V.: the German-Israeli Foundation (GIF No. I-1500-303.7/2019); A. Be, a. Bo, A. D.L., G. I. and D. R. received funding from the European Research Council (ERC No.852016); A. Be. and J. B.-C.: the University of Naples-Compagnia di San Paolo grant STAR-2016; L. Cs.,T.S, F.Zs, Gy.Gy: the Hungarian National Research Development and Innovation Office NKFIH (contract numbers PD129060, K120666 and K134197); T.S.: the János Bolyai research fellowship of the Hungarian Academy of Sciences and ÚNKP-20-5-DE-297 New National Excellence Program of the Ministry of Human Capacities of Hungary;D.B.: DFG (BE 4100/4-1); M. A., C.G.B., T.C., T. D.: STFC-UK. R.D. and D.P. acknowledge funding from the Italian Ministry of Education, University and Research (MIUR) through the ‘Dipartimenti di eccellenza’ project Science of the Universe. A. Be. is grateful to Richard J. deBoer for discussions about the R-matrix analysis.
\end{acknowledgments}

%\bibliographystyle{apsrev}
%\bibliography{c13an-letter}% Produces the bibliography via BibTeX.


\end{document}